\documentclass[reprint,twocolumn,pre,superscriptaddress,amsmath,amssymb,aps]{revtex4-2}

\usepackage{natbib}
\usepackage{graphicx,color,rotating}
\usepackage[latin1]{inputenc}
\usepackage{textcomp}
\usepackage{dcolumn}
\newcolumntype{d}[1]{D..{#1}}
\newcommand\mc[1]{\multicolumn{1}{c}{#1}} 
\newcommand\mrr[1]{\multicolumn{1}{r}{#1\;\;}} 
\newcommand\mlb[1]{\multicolumn{1}{l|}{#1}} 
\usepackage{bm}     
\usepackage{upgreek}
\usepackage{subdepth}  			
\usepackage{epstopdf}   		
\usepackage[latin1]{inputenc}
\usepackage{amsmath}
\usepackage{amssymb}
\usepackage{amsfonts}
\usepackage{array}
\newcolumntype{L}[1]{>{\raggedright\let\newline\\\arraybackslash\hspace{0pt}}m{#1}}
\newcolumntype{C}[1]{>{\centering\let\newline\\\arraybackslash\hspace{0pt}}m{#1}}
\newcolumntype{R}[1]{>{\raggedleft\let\newline\\\arraybackslash\hspace{0pt}}m{#1}}

\usepackage{hyperref}					
\usepackage[table,xcdraw]{xcolor}					
\hypersetup{			     			
    colorlinks,
    allcolors={blue}
}
\newcommand{\qmarks}[1]{{``#1''}}

\newcommand{\upspace}{\rule{0ex}{2.5ex}}

\newcommand{\mr}[1]{\ensuremath{\mathrm{#1}}}
\newcommand{\myvec}[1]{\bm{#1}}
\newcommand{\ee}{\mathrm{e}}
\newcommand{\ii}{\mathrm{i}}
\newcommand{\dm}{\mathrm{d}}

\newcommand{\pare}[1]{\left( #1 \right)}

\DeclareMathOperator{\re}{Re}
\DeclareMathOperator{\im}{Im}

\newcommand{\iot}{{\ii\omega t}}

\newcommand{\ve}{\varepsilon}
\newcommand{\vebf}{\bm{\varepsilon}}

\newcommand{\pp}{\partial}

\newcommand{\nablabf}{\boldsymbol{\nabla}}

\newcommand{\divop}{\nablabf\cdot}

\newcommand{\etal}{\textit{et~al.}}

\newcommand{\CCC}{\myvec{C}}

\newcommand{\DDD}{\myvec{D}}

\newcommand{\EEE}{\myvec{E}}

\newcommand{\eee}{\myvec{e}}

\newcommand{\JJJ}{\myvec{J}}

\newcommand{\KKK}{\myvec{K}}

\newcommand{\MMM}{\myvec{M}}

\newcommand{\nnn}{\myvec{n}}

\newcommand{\PPP}{\myvec{P}}

\newcommand{\rrr}{\myvec{r}}

\newcommand{\sss}{\myvec{s}}
\newcommand{\SSS}{\myvec{S}}

\newcommand{\uuu}{\myvec{u}}

\newcommand{\calC}{\mathcal{C}}

\newcommand{\calS}{\mathcal{S}}

\newcommand{\aL}{\alpha_\mr{lo}}
\newcommand{\aT}{\alpha_\mr{tr}}

\newcommand{\etab}{\eta^\mathrm{b}}

\newcommand{\SICel}{^\circ\!\textrm{C}}
\newcommand{\SIcm}{\textrm{cm}}
\newcommand{\SIum}{\upmu\textrm{m}}

\newcommand{\SIHz}{\textrm{Hz}}

\newcommand{\SIMHz}{\textrm{MHz}}

\newcommand{\SIkgm}{\textrm{kg}\:\textrm{m$^{-3}$}}

\newcommand{\SIm}{\textrm{m}}

\newcommand{\SImum}{\textrm{\textmu{}m}}

\newcommand{\SINp}{\textrm{Np}}

\newcommand{\SIGPa}{\textrm{GPa}}

\newcommand{\SIs}{\textrm{s}}

\newcommand{\SImus}{\textrm{\textmu{}s}}

\newcommand{\SIV}{\textrm{V}}

\newcommand{\SImW}{\textrm{mW}}
\newcommand{\nn}{\nonumber}
\newcommand{\beq}[1]{\begin{equation} \eqlab{#1}}
\newcommand{\eeq}{\end{equation}}
\newcommand{\bsub}{\begin{subequations}}
\newcommand{\esub}{\end{subequations}}
\def\bal#1\eal{\begin{align}#1\end{align}}
\def\balat#1#2\ealat{\begin{alignat}{#1} #2 \end{alignat}}
\def\bsubal#1 #2\esubal{\bsuba{#1}\begin{align}#2\end{align} \esuba}     
\def\bsubalat#1#2#3\esubalat{\bsuba{#1} \begin{alignat}{#2} #3 \end{alignat} \esuba}
\newcommand{\bsuba}[1]{\bsub \eqlab{#1}}
\newcommand{\esuba}{\esub}

\newcommand{\eqlab}[1]{\label{eq:#1}}
\renewcommand{\eqref}[1]{Eq.~(\ref{eq:#1})}

\newcommand{\eqsref}[2]{Eqs.~(\ref{eq:#1}) and~(\ref{eq:#2})}
\newcommand{\eqsnoref}[2]{(\ref{eq:#1}) and~(\ref{eq:#2})}

\newcommand{\figref}[1]{Fig.~\ref{fig:#1}}

\newcommand{\figsref}[2]{Figs.~\ref{fig:#1} and~\ref{fig:#2}}
\newcommand{\figlab}[1]{\label{fig:#1}}

\newcommand{\secref}[1]{Section~\ref{sec:#1}}

\newcommand{\secsref}[2]{Sections~\ref{sec:#1} and~\ref{sec:#2}}
\newcommand{\seclab}[1]{\label{sec:#1}}
\newcommand{\tabref}[1]{Table~\ref{tab:#1}}
\newcommand{\tabnoref}[1]{\ref{tab:#1}}
\newcommand{\tabsref}[2]{Tables~\ref{tab:#1} and~\ref{tab:#2}}
\newcommand{\tablab}[1]{\label{tab:#1}}

\newcommand{\sigmabf}{\bm{\sigma}}

\newcommand{\kL}{k_\mathrm{lo}}
\newcommand{\kT}{k_\mathrm{tr}}

\newcommand{\cL}{c_\mathrm{lo}}
\newcommand{\cT}{c_\mathrm{tr}}

\definecolor{darkgreen}{rgb}{0.00, 0.50, 0.00}
\definecolor{DARKGREEN}{rgb}{0.00, 0.50, 0.00}
\definecolor{RED}{rgb}{1.00, 0.00, 0.00}
\definecolor{GREEN}{rgb}{0.00, 1.00, 0.00}
\definecolor{BLUE}{rgb}{0.00, 0.00, 1.00}
\definecolor{MAGENTA}{rgb}{1.00, 0.00, 1.00}

\begin{document}

\title{Determination of the complex-valued elastic moduli of polymers by electrical impedance spectroscopy for ultrasound applications}

\author{William N. Bod\'{e}}
\email{winabo@dtu.dk}
\thanks{Equal share first-author}
\affiliation{Department of Physics, Technical University of Denmark,
DTU Physics Building 309, DK-2800 Kongens Lyngby, Denmark}

\author{Fabian Lickert}
\email{fabianl@dtu.dk}
\thanks{Equal share first-author}
\affiliation{Department of Physics, Technical University of Denmark,
DTU Physics Building 309, DK-2800 Kongens Lyngby, Denmark}

\author{Per Augustsson}
\email{per.augustsson@bme.lth.se}
\affiliation{Department of Biomedical Engineering, Lund University, Ole R\"{o}mers v\"{a}g 3, 22363 Lund, Sweden}

\author{Henrik Bruus}
\email{bruus@fysik.dtu.dk}
\affiliation{Department of Physics, Technical University of Denmark,
DTU Physics Building 309, DK-2800 Kongens Lyngby, Denmark}

\date{30 September 2022}

\begin{abstract}
A method is presented for the determination of complex-valued compression and shear elastic moduli of polymers for ultrasound applications. The resulting values, which are scarcely reported in the literature, are found with uncertainties typically around 1\% (real part) and 6\% (imaginary part). The method involves a setup consisting of a cm-radius, mm-thick polymer ring glued concentrically to a disk-shaped piezoelectric transducer. The ultrasound electrical impedance spectrum  of the transducer is computed numerically  and fitted to measured values as an inverse problem in a wide frequency range, typically from 500~Hz to 5~MHz, both on and off resonance. The method was validated experimentally by ultrasonic through-transmission around 1.9~MHz. The method is low cost, not limited to specific geometries and crystal symmetries,  and, given the developed software, easy to execute.  The method has no obvious frequency limitations before severe attenuation sets in above 100~MHz.
\end{abstract}

\maketitle

\section{Introduction}
Numerical simulations play an important role when optimizing and predicting piezoelectric device performance in applications including ultrasonic cleaning  \cite{Bretz2005}, energy harvesting \cite{Todaro2017}, inkjet printing \cite{Singh2010}, and acoustofluidics \cite{Bode2020, Lickert2021}. To perform precise, accurate, and predictive simulations, well-characterized material parameters such as the complex-valued elastic moduli are required. Whereas material databases exist \citep{MatWeb}, and manufactures may provide some of the required parameters, it is often not sufficient when attempting to perform reliable simulations and predictions. Polymers are in this regard a particularly challenging class of materials, since the elastic moduli of a given polymer may depend on unspecified parameters such as the distribution of polymer chain lengths and fabrication processes.

There exist a range of techniques to characterize an unknown material or substance mechanically. Dynamic techniques such as resonant ultrasound spectroscopy \cite{Migliori1993}, transmission techniques \cite{Wang1999, Xu2020}, impulse excitation \cite{Roebben1997}, laser vibrometry and triangulation  \cite{Willis2001, Ilg2012}, as well as static techniques, such as four-point bending, are widely used in various industries \cite{Radovic2004}. Those methods however often rely on a few mechanical eigenmodes or resonance frequencies of the material under study, a broad frequency spectrum due to a narrow pulse in the time domain, or even static or low-frequency measurements. Applications requiring actuation frequencies in the MHz-range however require material properties that were measured in similar frequency intervals for an accurate description of the system.

In this work we aim to extend the field of ultrasound spectroscopy \cite{Ilg2012, Rupitsch2009} by utilizing an electrical impedance spectrum spanning a frequency range of several MHz to obtain a full set of complex-valued elastic moduli of polymers. With this technique, labeled ultrasound electrical impedance spectroscopy (UEIS), a piezoelectric disk, driving vibrations in an attached polymer ring, is used to characterize the complex-valued elastic compressional and shear moduli of the polymer ring. Similar techniques have been used in the past to fit piezoelectric material parameters by an inverse problem and numerical optimization procedures on a free oscillating piezoelectric transducer \cite{Wang1999, Plesek2004, Rupitsch2009, Perez2010, Ilg2012, Perez2014, Kiyono2016}. Here, the same principles are used to fit elastic material parameters. From the UEIS spectrum of a mass-loaded transducer, an inverse problem is constructed to deduce the elastic moduli of the mass load. The method is similar to those of Refs.~\cite{Maynard1996, Ilg2012, Steckel2021}, but by including an automated whole-spectrum fit and complex parameter values, it extends the previous method as suggested in the conclusion of Ref.~\cite{Steckel2021}. Instead of a thin-film transducer and manual fitting of few selected resonance peaks in the impedance spectrum, the UEIS method makes use of several hundred impedance values measured on a mechanically-loaded bulk transducer in the frequency range from 500 Hz to 5 MHz, to extract both real and imaginary parts of the complex-valued elastic moduli, and not just the real parts obtained in Ref.~\cite{Steckel2021}. The UEIS technique enables low-cost and in-situ measurements of elastic moduli over a wide frequency range from low kHz to several MHz. It is easy to execute, requiring only a disk-shaped piezoelectric transducer, a ring of the unknown polymer sample, an impedance analyzer, and the developed fitting software.

The paper is organized as follows. In \secref{theory} a brief overview of the relevant theory is given, before in \secref{methodology} the experimental and numerical methodology of the UEIS technique is described in detail for polymer, glue, and transducer. In \secref{UTTresults} we provide validation data based on ultrasonic-through-transmission (UTT) measurements, before we in \secref{UEISresults} present the main results of the UEIS method in terms of the complex-valued electromechanical parameters of the unloaded piezoelectric transducer and the complex-valued elastic moduli of the UV-cured glue and the polymer ring.  We conclude in \secref{conclusion}.
\\[-6mm]

\section{Theoretical Background}
\seclab{theory}
We follow Ref.~\cite{Skov2019} and describe isotropic polymers using the standard linear theory of elastic solids in the Voigt notation, in terms of the displacement vector $\uuu$ of a given material point away from its equilibrium position, and the $1\times 6$  strain $\sss$ and stress $\sigmabf$ column vectors with the $6\times 1$ transposed row vectors $\sss^\textsf{T} = (\pp_x u_x,\pp_y u_y,\pp_z u_z, \pp_y u_z +\pp_z u_y, \pp_x u_z +\pp_z u_x, \pp_x u_y +\pp_y u_x)$ and  $\sigmabf^\textsf{T} = (\sigma_{xx}, \sigma_{yy}, \sigma_{zz}, \sigma_{yz}, \sigma_{xz}, \sigma_{xy})$, respectively. Representing the elastic moduli $C_{ik}$ by the tensor $\CCC$, the constitutive equation for an elastic solid in the $\infty m m$-symmetry class is \cite{Ikeda1996},
 \bsubal{ConstitPolymer}
 \eqlab{sigmaPolymer}
 \sigmabf &= \CCC\cdot\sss,
 \\
 \eqlab{CPolymer}
 \CCC & = \pare{
 \begin{array}{cccccc}
 C_{11} & C_{12} & C_{13} &0 &0 &0  \\
 C_{12} & C_{11} & C_{13} &0 &0 &0  \\
 C_{13} & C_{13} & C_{33}  &0 &0 &0  \\
 0 & 0 & 0 & C_{44} & 0 & 0 \\
 0 & 0 & 0  & 0 & C_{44} & 0 \\
 0 & 0 & 0 & 0 & 0 & C_{66} \\
 \end{array}
 }.
 \esubal
For an isotropic polymer $C_{33} =  C_{11}$,  $C_{66} =  C_{44}$, and
$C_{13} = C_{12} = C_{11}-2C_{44}$, so here, $\CCC$ is given only by the two complex-valued elastic moduli $C_{11}$ and $C_{44}$, each with a real and imaginary part, $C_{ik} = C'_{ik} + \ii C''_{ik}$, relating to the propagation and attenuation of sound waves, respectively. Many amorphous polymers, such as the injection-molded PMMA in this work, are isotropic, but if not, such as semi-crystalline polymers \cite{Michler2016}, a $\CCC$ tensor with the appropriate lower symmetry must be used. Since only positive power dissipation is allowed, the elastic moduli are restricted by the constraint that the matrix $\im(-\CCC)$ must be positive definite \cite{Holland1967}.

We also model the piezoelectric lead-zirconate-titanate (PZT) transducer in the $\infty m m $-symmetry class \cite{Ikeda1996}, again following the notation of Ref.~\cite{Skov2019}. Here, $\uuu$, $\SSS$, $\sigmabf$, and $\CCC$ are supplemented by the electric potential $\varphi$, the electric field $\EEE = -\nablabf\varphi$, the dielectric tensor $\vebf$, the electric displacement field $\DDD = \vebf\cdot\EEE$, and the piezoelectric  $3\times6$ coupling tensor $\eee$. The constitutive equation becomes,
 \bsubal{ConstitPZT}
 \eqlab{sigmaDPZT}
 &\pare{\begin{array}{c} \sigmabf \\ \DDD \end{array}}
 =
 \pare{\begin{array}{cc}\CCC & -\eee^\textsf{T}  \\ \eee & \vebf \end{array}}
 \,
 \pare{\begin{array}{c} \SSS\\ \EEE \end{array}}
 =
 \MMM\cdot
 \pare{\begin{array}{c} \SSS\\ \EEE \end{array}}
 \\
 \eqlab{ePZT}
 &\eee  = \pare{
 \begin{array}{cccccc}
 0 &0 &0 &0 &e_{15} &0  \\
 0 &0 &0 &e_{15} &0 &0  \\
 e_{31} &e_{31} &e_{33} &0 &0 &0
 \end{array}
 }\!,
 \;
 \vebf  = \pare{
 \begin{array}{ccc}
 \ve_{11} &0 &0\\
 0 &\ve_{11}  &0\\
 0 &0 &\ve_{33}
 \end{array}
 }\!.
 \esubal
In the $\infty m m $-symmetry class, $C_{66}=\frac{1}{2}(C_{11}-C_{12})$, so the coupling tensor $\MMM$ is given by the five complex-valued elastic moduli $C_{11}$, $C_{12}$, $C_{13}$, $C_{33}$, and $C_{44}$, with $C_{ik} =  C'_{ik} + \ii C''_{ik}$, the two complex-valued dielectric constants $\ve_{11}$ and $\ve_{33}$ with $\ve_{ik} = \ve'_{ik} + \ii \ve''_{ik}$, and the three real-valued piezo-coupling constants $e_{31}$, $e_{33}$, and $e_{15}$ with $e_{ik} = e'_{ik} + 0\ii$. Since only positive power dissipation is allowed, the coupling constants are restricted by the following constraint on the matrix $\KKK$ \cite{Holland1967},
 \beq{Kconstraint}
 \KKK =
 \im\pare{\begin{array}{cc} -\CCC & \eee^\textsf{T}  \\ \eee & \vebf \end{array}}\;
 \text{ must be positive definite}.
 \eeq

We limit our analysis of the linear system to the time-harmonic response for a given angular frequency $\omega = 2\pi f$, where $f$ is the excitation frequency of the system. Thus, any physical field $F_\mr{phys}(\rrr,t)$ is given by a complex-valued amplitude $F(\rrr)$ as $F_\mr{phys}(\rrr,t) = \re\big[F(\rrr)\:\ee^{-\iot}\big]$, and we need only to compute $F(\rrr)$. In our model of a polymer sample mounted on a PZT transducer having a bottom and top electrode, the system is excited by the excitation voltage $\varphi_\mr{top}^\mr{phys} = \re\big[\varphi_\mr{top}\: \ee^{-\iot}\big]$ as follows,
 \beq{BC_elec}
 \varphi_\mr{bot} = 0~\SIV \quad \text{ and } \quad \varphi_\mr{top} = 1~\SIV.
 \eeq
By introducing the density $\rho$ as an additional material parameter, the governing equations for the time-harmonic displacement field $\uuu$ in the polymer and in the PZT and for the quasi-electrostatic potential $\varphi$ in the non-magnetic PZT without free charges, become
 \bal
 \divop \sigmabf = -\rho \omega^2 \uuu \quad \text{ and } \quad
 \divop \DDD = 0.
 \eal
We neglect the effect of gravity in this formulation, as it only leads to a minor deformation of the geometry. The stress- and charge-free boundary conditions are imposed on free surfaces
 \beq{BCs}
 \sigmabf\cdot\nnn = \bm{0} \quad \text{ and } \quad  \nnn\cdot \DDD = 0.
 \eeq
The current density $\JJJ$ in the PZT transducer is given by the polarization $\PPP$ as
 \bal
 \JJJ = \partial_t \PPP = -\ii \omega (\DDD-\ve_0 \EEE).
 \eal
Consequently, the electrical impedance $Z$ central to the UEIS method can be computed via the flux integral of $\JJJ$ through the surface $\partial \Omega_\mr{top}$ with surface normal $\nnn$ as,
 \beq{Zdef}
 Z = \frac{\varphi_\mr{top}-\varphi_\mr{bot}}{I},\quad \text{ with }\quad
 I = \int_{\partial \Omega_\mr{top}} \nnn\cdot \JJJ\: \dm A.
 \eeq

\section{Methodology}
\seclab{methodology}

The ultimate goal is to develop and test a method for determination of the complex-valued elastic moduli of polymers. However, to achieve an accuracy level of about 1-5\%, we need also to determine the mechanical and electromechanical parameters of the piezoelectric transducer as well as the elastic moduli of the glue used to mount the polymer sample on the transducer.

\begin{table}[b]
\centering
\caption{\tablab{SampleDim} Measured thickness (TH) and outer diameter (OD) of the Pz27 disks (named Pz27-TH-OD). Measured TH, OD, and inner diameter (ID) of the rings made by the glue NOA 86H (named NOA86H-TH-OD) and of the polymer PMMA (named PMMA-TH-OD). Letters A, B, C are used as labels for samples with the same nominal dimensions. The precision is given as the standard deviation from six measurements.
}
\begin{ruledtabular}
\begin{tabular}{l|d{1.6}d{2.6}d{2.5}}
Sample     & \mc{TH}   & \mc{OD}    & \mc{ID}\\
           & \mc{(mm)} & \mc{(mm)}  & \mc{(mm)}\\
\hline
Pz27-0.5-6.35-A & 0.510(1)  & 6.594(5)   & \mc{--}\upspace\\
Pz27-0.5-6.35-B & 0.502(1)  & 6.587(2)   & \mc{--}\\
Pz27-0.5-10-A & 0.500(1)  & 10.037(7)  & \mc{--}\\
Pz27-0.5-10-B & 0.492(2)  & 10.039(3)  & \mc{--}\\
Pz27-0.5-10-C & 0.505(2)  & 10.039(5)  & \mc{--}\\
NOA86H-1.4-20 & 1.5(2)    & 19.22(6)  & 1.90(2) \\
PMMA-1.4-20-A & 1.428(2)  & 19.981(5) & 1.98(1) \\
PMMA-1.4-20-B & 1.440(4)  & 20.017(8) & 1.94(1) \\
PMMA-1.4-25-A & 1.427(5)  & 24.98(1)  & 1.92(1) \\
PMMA-1.4-25-B & 1.437(5)  & 24.94(1)  & 1.94(2) \\
\end{tabular}
\end{ruledtabular}
\end{table}

\subsection{Experimental procedure}
\seclab{experimental_procedure}
Measurements of the electrical impedance spectrum $Z_\mr{exp}(f)$ were performed using a Vector Network Analyzer Bode 100 (OMICRON electronics GmbH, Klaus, Austria) in a linear interval with 10.000 steps between 500~Hz and 5~MHz, each frequency sweep taking less than 4~minutes. In a first step, the electrical impedance of an unloaded Pz27 disk (Meggitt A/S, Kvistgaard, Denmark) was characterized. The top and bottom electrode of the piezoelectric disk were contacted through spring-loaded pins in a custom-made sample holder, minimizing the contact force and area to a point in the center of the disk. In the following step, using a thin layer of the UV-curable glue NOA 86H (Norland Products, Jamesburg (NJ), USA), a polymer ring of known dimensions was glued on top of a Pz27 disk. This ring was made from either the NOA 86H glue itself or from polymethyl methacrylate (PMMA Diakon TD525, Lucite International, Rotterdam, Netherlands). We aimed at making the glue layer as thin ($15$-$25~\SImum$) and uniform as possible  to reduce its influence on the combined system, as studied by Bod\'e \etal~\cite{Bode2021}. This was achieved by gently squeezing the system for a few seconds after assembly, just before curing the glue by UV illumination. The glue was cured at a UV-intensity of $15~\SImW/\SIcm^2$ at 365~nm for 167~s to achieve strong bonding between the transducer disk and the polymer ring. The glue NOA 86H was selected after performing experiments with several different adhesives, as it enables good adhesion between the PZT and the polymer and allows for good acoustic coupling due to acrylic-like properties in the cured state with an attenuation comparable to that found for the polymer ring. After curing, the electrical impedance of the polymer-loaded transducer was measured. The small hole of the polymer ring allows contacting the transducer disk using the above-mentioned spring-loaded pins. The average of three impedance measurements, taking less than 12~minutes to obtain, was used both for the unloaded and loaded case.

The diameter and thickness of the polymer ring and the Pz27 disk were measured before assembling the system using an electronic micrometer with an accuracy of $\pm 4~\SImum$. The glue-layer thickness was obtained as the measured total thickness of the assembled system minus the sum of the individual thicknesses of the Pz27 disk and the polymer ring. The impedance measurements were performed at $24~\SICel$ using a combination of two different nominal transducer dimensions (diameter 6.35~mm and 10~mm, thickness 0.5~mm) and two different nominal polymer ring dimensions (diameter 20~mm and 25~mm, thickness 1.4~mm), yielding four transducer-polymer systems with the dimensions listed in \tabsref{SampleDim}{config}.

\subsection{Numerical model}
The weak formulation of the finite element method (FEM) is used to implement the governing equations in the software COMSOL Multiphysics \cite{Comsol55} to simulate the electrical impedance spectrum $Z_\mr{sim}(f)$ unloaded or loaded PZT transducer. In particular, we use the \textit{weak form PDE} interface as described in our previous work \cite{Skov2019, Lickert2021, Bode2021}. The simulations are computed on a workstation with a 12-core, 3.5-GHz central processing unit and 128 GB random access memory. Third-order Lagrange polynomials are used as test functions for both $\uuu$ and $\varphi$. The model consists of three domains: a piezoelectric disk, a glue layer, and a polymer ring. Given the cylindrical geometry of the assembled stack and the axisymmetric structure of the coupling tensors $\CCC$ and $\MMM$ in \eqsref{CPolymer}{sigmaDPZT}, the system can be reduced to an axisymmetric model as shown in Ref.~\cite{Steckel2021c} and illustrated in \figref{model}. This axisymmetrization reduces the computational time substantially. A suitable mesh element size is found by the mesh convergence study presented in Sec.~S1 of the Supplemental Material~\footnote{See Supplemental Material at [{URL}] for details on the mesh convergence analysis, a study on the sensitivity versus elastic moduli and frequency range, sample MATLAB and COMSOL scripts for the UEIS fitting procedure, the corresponding data for the impedance spectra, and the validation of the UEIS method by UTT and laser-Doppler vibrometry, which includes Refs.~[16, 29].},
where in addition in Sec.~S2, a COMSOL sample script is presented.

\begin{table}[t]
\centering
\caption{\tablab{config} The measured glue-layer thickness of the five studied PMMA-ring-on-Pz27-disk configurations.}
\begin{ruledtabular}
\begin{tabular}{lc|lc|r}
Pz27 disk    & & Polymer ring & & \mc{Glue layer} \\
\hline
\upspace Pz27-0.5-6.35-A & & PMMA-1.4-20-A & & 15~$\SImum$ \\
Pz27-0.5-6.35-B & & PMMA-1.4-25-A & & 24~$\SImum$ \\
Pz27-0.5-10-A & & PMMA-1.4-25-B & & 21~$\SImum$ \\
Pz27-0.5-10-B & & PMMA-1.4-20-B & & 12~$\SImum$ \\
Pz27-0.5-10-C & & NOA86H-1.4-20 & & 15~$\SImum$ \\
\end{tabular} \end{ruledtabular}
\end{table}

Using the \qmarks{\texttt{LiveLink for MATLAB}}-interface provided by COMSOL, the MATLAB optimization procedures \texttt{fminsearchbnd} and \texttt{patternsearch} are used to fit the material parameters such that $Z_\mr{sim}(f)$ is as close to $Z_\mr{exp}(f)$ as possible. The \texttt{fminsearchbnd} algorithm \cite{fminsearchbnd} allows a bounded search in parameter space. The \texttt{patternsearch} algorithm (part of the \texttt{Global Optimization Toolbox}) makes twice as many function evaluations, but it covers a larger region in parameter space and is better to locate the global minimum for poor initial values. Both algorithms use a gradient-free direct search and are therefore well suited for non-smooth numerical optimization procedures. The algorithms require three inputs: (i) initial values, (ii) upper and lower bounds, and (iii) a cost function to minimize. Based on the measured and simulated electrical impedance values $Z_\mr{exp}(f_i)$ and $Z_\mr{sim}(f_i)$ obtained at $\sim 500$ frequencies $f_i$, we define the cost function $\mathcal{C}$ as
 \beq{calCdef}
 \mathcal{C} = \sqrt{{\sum_i \big[\log_{10}\!\big(|Z_\mr{exp}(f_i)|\big)
    - \log_ {10}\!\big(|Z_\mr{sim}(f_i)|\big)\big]^2}}.
 \eeq
Here, we use the logarithm, because $Z(f_i)$, having many peaks, varies by orders of magnitude as a function of $f_i$.

\begin{figure}[t]
    \centering
    \includegraphics{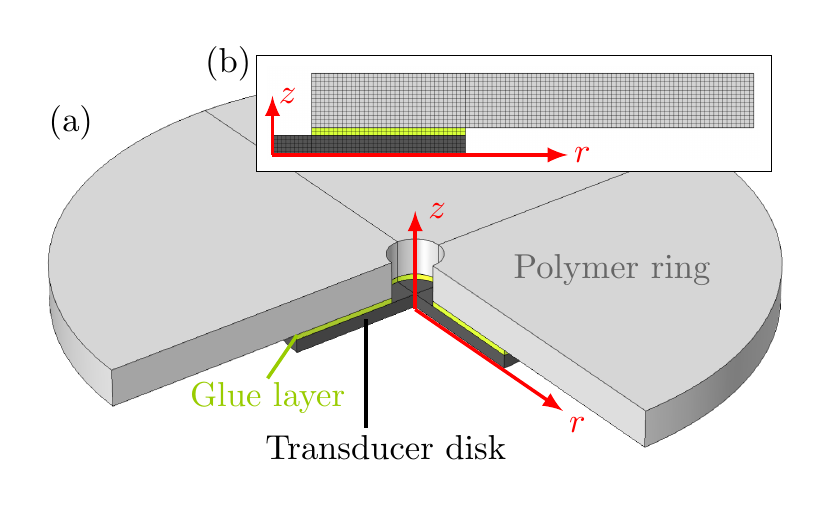}
    \caption{\figlab{model}
    (a) A 3D sketch of the system consisting of a polymer ring (light gray) glued (green) to a transducer disk (dark gray) whit a quarter cut away for visibility. (b) The inset shows the 2D axisymmetric domain in the $r$-$z$ plane used for numerical simulations. The structured mesh is the one used at 5~MHz.}
\end{figure}

\subsection{Sensitivity analysis}

To enhance the performance of our fitting procedure, we group the parameters into sets of similar sensitivity based on the following sensitivity analysis of the cost function $\calC$ on each of the sixteen material parameters $(p_1, p_2, \ldots p_{16})^\mr{pz27} = (C_{11}'$, $C_{12}'$, $C_{13}'$, $C_{33}'$, $C_{44}'$, $\ve_{11}'$, $\ve_{33}'$, $e_{31}'$, $e_{33}'$, $e_{15}'$, $C''_{11}$, $C''_{12}$, $C''_{13}$, $C''_{33}$, $C''_{44}$, $\ve''_{33})$ for the Pz27 disk and on the four polymer parameters $(p_1, \ldots p_4)^\mr{polym} = (C_{11}'$, $C_{44}'$, $C''_{11}$, $C''_{44})$. The sensitivity analysis is performed in the frequency range from 500~Hz to 5~MHz, with the initial value $p_i^0$ taken from literature for a given parameter $p_i$, and therefore the individual sensitivity values represent averages over the entire frequency range. A more detailed study of the frequency dependency of the sensitivity is shown in Sec. S2 of the Supplemental Material \cite{Note1}. We use a discrete approximation of the relative sensitivity $\calS(p_i) = p_i \pp_{p_i}\calC(p_i)$ of $\calC(p_i)$ based on a $\pm10$\% variation of $p_i$ around $p_i^0$, while keeping the remaining parameters fixed at $p_j^0$,
 \beq{SiDef}
 \calS(p_i) = \frac{\calC(1.1p_i)-\calC(0.9p_i)}{1.1-0.9},
 \;\; p_j = p_j^0 \text{ for }\; j\neq i.
 \eeq
The obtained sensitivities $\calS$ for the $p_i^\mr{pz27}$ and $p_i^\mr{pmma}$ parameters for Pz27 and PMMA, respectively, are shown in \figref{sensitivity}. The Pz27 parameters are classified in three groups of high $\calS(p_i)>1$, medium $0.1<\calS(p_i)<1$, and low $\calS(p_i)<0.1$ sensitivity, respectively, and as described in the following section, a robust fitting is obtained by fitting the parameters group by group sequentially in descending order from high to low sensitivity. Since all four PMMA parameters have a medium-to-high sensitivity we fit them simultaneously in a single, undivided group.

\begin{figure}[t]
 \centering
 \includegraphics{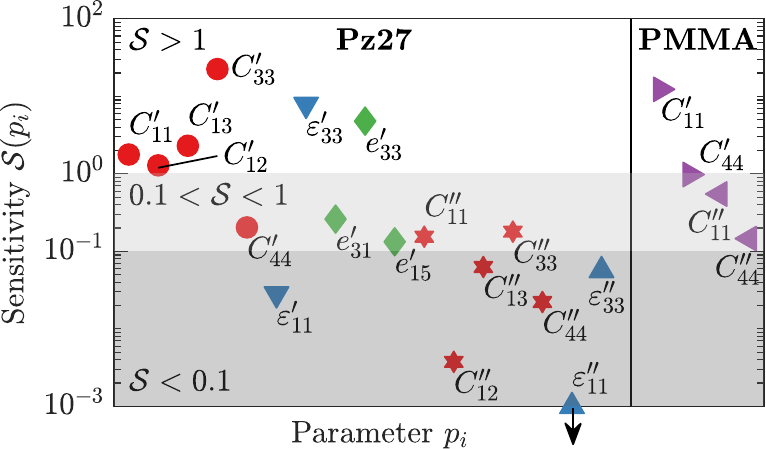}
 \caption{\figlab{sensitivity}
The relative cost function sensitivity $\calS(p_i)$ for the 17 piezoelectric material parameters $p_i$ obtained as an average from four Pz27 disks in the frequency interval 500~Hz$-$5~MHz is shown in the left side of the figure. $\calS(p_i)$ for the four PMMA parameters, calculated from the average of four Pz27--PMMA--systems, are shown on the right. Corresponding real and imaginary parts are visualized in the same color, and the regions of high ($\calS(p_i)>1$), medium ($0.1<\calS(p_i)<1$) and low ($\calS(p_i)<0.1$) sensitivity are highlighted by gray shadows. The sensitivity of the parameter $\ve''_{11}$ is close to zero, as indicated by a black arrow.}
\end{figure}

\subsection{The UEIS fitting procedure}
\seclab{FitProc}
The first step in the UEIS fitting procedure is to measure and simulate the electrical impedance $Z_\mr{exp}(f)$ and $Z_\mr{sim}(f)$, respectively, of an unloaded Pz27 transducer disk  and then  following Refs.~\cite{Perez2010, Perez2014, Kiyono2016} to fit the sixteen Pz27 parameters $p_i^\mr{pz27}$ in the form of an inverse problem by minimizing the cost function $\calC$. In the second step, a characterized Pz27 disk is loaded by gluing on a given polymer ring using the UV-curable glue NOA 86H. To characterize the glue, the first studied polymer ring is made by the glue itself, and $Z(f)$ is used to similarly fit the four glue parameters $p_i^\mr{glue}$. Subsequently, using the characterized glue, a PMMA ring is glued to a characterized Pz27 disk, and $Z(f)$ is used to similarly fit the four PMMA parameters $p_i^\mr{pmma}$. See the flow chart in \figref{fitting_proc}.

\begin{figure}[t]
\centering
\includegraphics[width=\columnwidth]{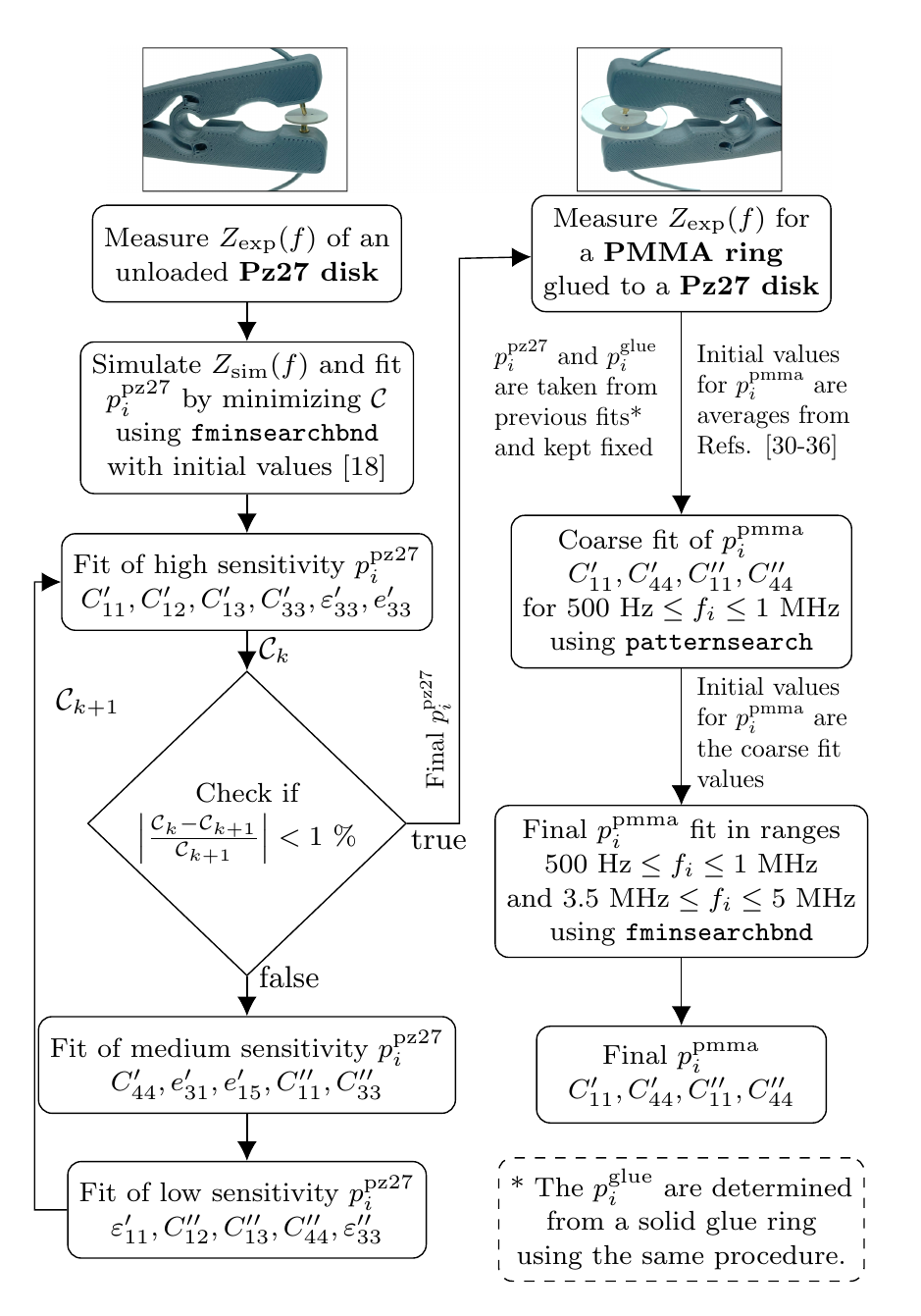}
\caption{\figlab{fitting_proc}
A flow chart of the steps in the fitting procedure to obtain complex-valued elastic moduli for a polymer sample. First, the electrical impedance spectrum $Z_\mr{exp}(f)$ of a Pz27 transducer disk is measured in the range $500~\SIHz < f_i < 5~\SIMHz$. Then, the Pz27 parameters $p_i^\mr{pz27}$ are fitted in the same frequency range with increments of $\Delta f = 10$~kHz based on their sensitivities in descending order, with initial values from Ref.~\cite{Kiyono2016}. Next step is the measurement of $Z_\mr{exp}(f)$ for the PMMA ring glued to the Pz27 disk in the range $500~\SIHz < f_i < 5~\SIMHz$. Then, coarse fitting of the $p_i^\mr{pmma}$ is performed in the range $500~\SIHz < f_i < 1~\SIMHz$ with $\Delta f = 2$~kHz stepping, taking the initial values to be the average values of Refs.~\cite{Hartmann1972, Christman1972, Sutherland1972, Sutherland1978, Carlson2003, Simon2019, Tran2016}. Lastly, a final fitting of $p_i^\mr{pmma}$ is done in the the range $3.5~\SIMHz < f_i < 5~\SIMHz$ combined with previous range using steps of $\Delta f = 10$~kHz. A MATLAB-COMSOL sample script for the PMMA fitting procedure is presented in Sec.~S2 of the Supplemental Material~\cite{Note1}.}
\end{figure}

For the unloaded Pz27 disk, the initial values of $p_i^\mr{pz27}$ are taken from Ref.~\cite{Kiyono2016}, and the $k$th iteration in the fit is divided into four sub-steps: (1) Fit the six parameters $C_{11}'$, $C_{12}'$, $C_{13}'$, $C_{33}'$, $\ve_{33}'$, and  $e_{33}'$ of highest sensitivity $\calS > 1$ using the \texttt{fminsearchbnd} algorithm in the range $500~\SIHz < f_i < 5~\SIMHz$ in increments of 10~kHz with the bounds set to $\pm30~\%$, while keeping the remaining eleven parameters fixed. (2) Check whether the cost function $\calC_{k}$ of iteration $k$ deviates less than 1~\% relative to $\calC_{k-1}$ (the fit is converged and $p_i^\mr{pz27}$ have been determined) or not (the fitting continues). (3) Similarly, fit the five parameters $C_{44}'$, $e_{31}'$, $e_{15}'$, $C''_{11}$, and $C''_{33}$ of medium sensitivity $0.1 < \calS < 1$. (4) Likewise, fit the last six parameters $\ve_{11}'$, $C''_{12}$, $C''_{13}$, $C''_{44}$, $\ve''_{11}$, and $\ve''_{33}$ of low sensitivity $\calS < 1$ and move on to iteration $k+1$. If during the fit a value of $p_i^\mr{pz27}$ is within 5~\% of the pre-defined bound, the latter is changed by 50~\%. Furthermore, for each evaluation of the cost function $\calC$, it is checked if $\KKK$ in \eqref{Kconstraint} is positive definite, and if not we set $\calC = \infty$.

For the glue ring, the initial values of the four parameters $p_i^\mr{glue}$ are $C_{11}^\mr{glue} = (4.7 -  0.47\ii)~\SIGPa$ and $C_{44}^\mr{glue} = (0.9 - 0.09\ii)~\SIGPa$ inferred from Young's modulus of Ref.~\cite{UVglueNOA86H}, the assumed value 0.38 of Poisson's ratio, and $C_{ik}'' = 0.1 C_{ik}'$. Moreover, the density of the glue ring is measured.  The fitting is divided into two sub-steps to increase robustness and speed: (1) A coarse fit of the four parameters $C_{11}'$, $C_{44}'$, $C_{11}''$, and $C_{44}''$ using the \texttt{patternsearch} algorithm in the limited range $500~\SIHz < f_i < 1~\SIMHz$ in increments of 2~kHz with the bounds set to be $\pm40~\%$ covering the typically observed range for polymers \cite{Carlson2003, Xu2020}. (2) A final fit of $C_{11}'$, $C_{44}'$, $C_{11}''$, and $C_{44}''$ using the \texttt{fminsearchbnd} algorithm in the combined ranges of  $500~\SIHz < f_i < 1~\SIMHz$ and $3.5~\SIMHz < f_i < 5~\SIMHz$ in increments of 2~kHz and 10~kHz, respectively, with the bounds set to $\pm40~\%$, and with the coarse-fit values used as initial values. If during the fit a value of $p_i^\mr{glue}$ is within 5~\% of the pre-defined bound, the bound is changed by 5~\%, see the Supplemental Material~\cite{Note1}. Furthermore, for each evaluation of the cost function $\calC$, it is checked if $\im(-\CCC)$ is positive definite, and if not we set $\calC = \infty$.

For the PMMA ring, the initial values of $p_i^\mr{pmma}$ are taken to be the average of the values reported in  Refs.~\cite{Hartmann1972, Christman1972, Sutherland1972, Sutherland1978, Carlson2003, Simon2019, Tran2016}. This average is used due to the lack of parameter values provided by the manufacturer of our selected PMMA polymer. Since this PMMA consists of a toughened acrylic compound, we expect that it deviates from standard PMMA grades. Therefore we chose to use the average literature values only as initial values in our fitting routine, and we refrain from comparing them with the resulting UEIS values. Otherwise, the fitting procedure for the PMMA ring is the same as the one for the glue ring.

Note that for the selected materials in the studied frequency range from 500~Hz to 5~MHz, and measured with relative accuracies from 1\% to 5\%, the experimental results in \secsref{UTTresults}{UEISresults} show that it is adequate to assume frequency-independent parameters $p_i^\mr{pz27}$, $p_i^\mr{glue}$ and $p_i^\mr{pmma}$. See further discussion in \secref{fittedPMMA}.

\section{Ultrasonic-through-transmission (UTT) validation data}
\seclab{UTTresults}
For the polymer PMMA, we have carried out ultrasonic-through-transmission (UTT) measurements \cite{Wang1999, Tran2016} to acquire data for experimental validation of the UEIS method. In UTT, a pulse, with center frequency $f_c$ and width $\Delta f$ in the frequency domain and width $\Delta t$ in the time domain, is transmitted through a polymer slab of thickness $d$ with its surface normal $\nnn$ tilted an angle $\theta_i$ relative to the incident pulse and emerged in water having the sound speed $c_0$. We have used $f_c = 1.90~\SIMHz$, $\Delta f \approx 1~\SIMHz$, and $\Delta t \approx 2~\SImus$. The UTT-method relies on the fact that at normal incidence only longitudinal waves are transmitted, whereas above a critical tilt angle $\theta_\mr{crit} = \sin^{-1}(c_0/\cL)$ only transverse waves are transmitted in samples with $\cL > c_0$. The longitudinal and transverse speed of sound, $\cL$ and $\cT$, and the corresponding attenuation coefficients, $\aL$ and $\aT$, of the slab can be determined based on the difference $\Delta \tau$ of arrival times, with and without the slab placed in the water,
 \bsubal{UTTdef_parameters}
 \eqlab{clo}
 \cL & = \Big[1+\frac{c_0}{d}\:\Delta \tau\Big]^{-1}\: c_0,
 \\
 \eqlab{ctr}
 \cT &=  \Big[\sin^2\theta_i + \big(\frac{c_0}{d}\:\Delta \tau
 + \cos\theta_i\big)^2\Big]^{-\frac12}\:c_0,
 \\
 \eqlab{alo}
 \aL &= \alpha_\mr{0}
 + \frac{1}{d}\ln{\!\bigg[\frac{T_\mr{lo} A_\mr{0}}{A_\mr{lo}}\bigg]},
 \\
 \eqlab{atr}
 \aT &= \alpha_\mr{0}\cos{(\theta_s-\theta_i)}
 + \frac{1}{d}\ln{\!\bigg[\frac{T_\mr{tr} A_\mr{0}}{A_\mr{lo}}\bigg]}\:\cos{\theta_s}.
 \esubal
Here,  $T_\mr{lo}$ and $T_\mr{tr}$ are the longitudinal and transverse transmission coefficients, $\theta_s$ is the refractive angle of the shear wave, $A_\mr{0}$ is the amplitude of the direct signal, and $A_\mr{lo}$ and $A_\mr{tr}$ are the longitudinal and transverse amplitudes of the transmitted signal after passing through the sample. Using the parameter values of water listed in Ref.~\cite{Muller2014}, the attenuation coefficient $\alpha_0$ of water is,
 \bal \eqlab{alphaWater}
 \alpha_0(T,f) &= 2 \pi^2 \bigg[\frac43 \eta_0(T) + \etab(T)\bigg]\frac{f^2}{\rho_0 c_0^3}
 \nn \\
 &\approx \bigg[1-0.0249\frac{T-25~\SICel}{1~\SICel}\bigg]\:
 \frac{0.0217 f^2}{1~\SIMHz^2}\frac{\SINp}{\SIm}.
 \eal

As we do not control the room temperature in our UEIS measurements, but simply monitor it with a $1~\SICel$ uncertainty, we have used the UTT experiments to determine the temperature dependence of the elastic moduli of our PMMA sample. To this end, the UTT tank was filled with warm water at temperature $T = 31~\SICel$. Then over a period of 6 hours, as the water steadily cooled to $T = 23~\SICel$, the  elastic moduli were measured at regular intervals, corresponding to steps in temperature of about $-0.5~\SICel$. As shown in Sec.~S4 of the Supplementary Material~\cite{Note1}, the resulting longitudinal and transverse speed of sound ($\cL$ and $\cT$) and attenuation coefficients ($\aL$ and $\aT$) of PMMA at the frequency $f_c = 1.90~\SIMHz$ are found to depend linearly on temperature $T$ (in $\SICel$) as,
 \bsubal{UTTresult}
 \cL^\mr{1.90\:MHz}(T) &=
 \bigg[-11(2)\:\frac{T}{1~\SICel} + 2743(46)\bigg]~\frac{\SIm}{\SIs},
 \\
 \cT^\mr{1.90\:MHz}(T) &=
 \bigg[-4.5(4)\:\frac{T}{1~\SICel} + 1267(9)\bigg]~\frac{\SIm}{\SIs},
 \\
 \aL^\mr{1.90\:MHz}(T) &=
 \bigg[1.7(4)\:\frac{T}{1~\SICel} + 25(10)\bigg]~\frac{\SINp}{\SIm},
 \\
 \aT^\mr{1.90\:MHz}(T) &=
 \bigg[7(2)\:\frac{T}{1~\SICel} + 146(43)\bigg]~\frac{\SINp}{\SIm}.
 \esubal
Here, the digits in the parentheses indicate $1\sigma$ uncertainties computed from on the sum-of-square differences between measured data and regression-line fits.\\[-7mm]

\section{Results of the UEIS method}
\seclab{UEISresults}

\subsection{UEIS-fitted material parameters for Pz27}
\seclab{fittedPz27}
First, we determine the sixteen material parameters $p_i^\mr{pz27}$ for the four unloaded Pz27 disks with nominal outer diameters 6.35~mm and 10.0~mm and the measured dimensions listed in \tabref{SampleDim}. Using the UEIS method described in \secref{methodology} involving the measured and fitted impedance spectra $Z_\mr{exp}(f)$ and $Z_\mr{sim}(f)$, we obtain the resulting parameters listed in \tabsref{Pz27_fit_real}{Pz27_fit_imag}. In \tabref{Pz27_fit_real} we compare the real part of the obtained UEIS parameters to those provided in the literature (lit, \cite{Kiyono2016}) and by the manufacturer (manf, \cite{Meggitt2021}). The relative difference $\Delta_\mr{UEIS}^\mr{lit}$ between UEIS and literature values is in the range $\sim$2-8~\%, whereas $\Delta_\mr{UEIS}^\mr{manf}$ is higher, typically in the range $\sim$5-40~\%. The deviations are overall significant compared to the relative standard deviation $\hat{\sigma}_\mr{UEIS} \lesssim 2$~\% of the mean of the UEIS values.

Similarly, in \tabref{Pz27_fit_imag} we compare the imaginary parts of the obtained UEIS parameters to those provided in the literature (lit, \cite{Kiyono2016}). Note that $\ve''_{11}$ is set to zero due to its low value and sensitivity, and that $e''_{mj} = 0$ by assumption. In general, the imaginary parts are more difficult to measure than the real parts, which is reflected in the high values of $\hat{\sigma}_\mr{UEIS}$ ($\lesssim$10~\%), $\Delta_\mr{UEIS}^\mr{lit}$ ($\sim$5-50~\%), but still with significant deviation between UEIS values and the values provided in the literature and by the manufacturer. The errors on the imaginary parts are about one order of magnitude larger than the errors on the real parts. This is in line with the previously found lower sensitivities of the former compared to the higher sensitivities of the latter shown in \figref{sensitivity}. Relative deviations of the initial values from the fitted values range from as little as 1.4~\% for $C''_{13}$ and up to 50~\% for $C''_{12}$ and above 200~\% for $\ve_{33}''$. Despite those deviations from the initial values, we find good convergence on the cost function and an excellent agreement between the measured and fitted impedance spectrum for the Pz27 disk. The uniqueness of the sixteen material parameters $p_i^\mr{pz27}$ is not guaranteed, but the simulated impedance spectrum fits the measured one, and thus they provide an adequate estimate for the subsequent determination of the polymer parameters $p_i^\mr{pmma}$. In \figref{UEIS_Pz27}, an example is shown of the measured UEIS spectrum and the resulting simulated UEIS spectrum for a Pz27 disk of diameter 10~mm and thickness 0.5~mm.

\begin{table*}[t]
\centering
\caption{\tablab{Pz27_fit_real}
The real part (single prime) of the material parameters $p_i^\mr{pz27}$ of the four Pz27 disks of \tabref{SampleDim} obtained by the UEIS method with initial values from the literature \cite{Kiyono2016}. The digits in brackets denote one standard deviation, $\hat{\sigma}_\mr{UEIS}$ is the relative standard deviation of the UEIS mean, and $\Delta_\mr{UEIS}^\mr{lit}$ and $\Delta_\mr{UEIS}^\mr{manf}$ is the relative deviation from UEIS of the values given by Ref.~\cite{Kiyono2016} (lit) and by the manufacturer (manf) \cite{Meggitt2021}, respectively. The density of Pz27 was measured to be $\rho_\mathrm{pz27}=7.7(1)$~g/cm$^3$.}
\begin{ruledtabular}
\begin{tabular}{c|cccccccccc}
Pz27 disk                                                     & $C'_{11}$  & $C'_{12}$  & $C'_{13}$  & $C'_{33}$& $C'_{44}$ & $\ve'_{11}$  & $\ve'_{33}$ & $e'_{31}$ & $e'_{33}$& $e'_{15}$  \\
               & (GPa)      & (GPa)      & (GPa) & (GPa)      & (GPa)      & ($\ve_0$) & ($\ve_0$)  & (C/m$^2$)  &(C/m$^2$) & (C/m$^2$)     \\
\hline
\mlb{Pz27-0.5-6.35 \hfill  (A)} \upspace & $127$        & $80.1$        & $82.3$        & $121$  & $20.3$   & $963$    & $780$               & $-5.3$        & $16.4$  & $11.8$      \\
\mlb{Pz27-0.5-6.35 \hfill (B)} & $126$        & $78.8$        & $81.4$        & $122$  & $20.7$   & $957$    & $753$               & $-5.2$        & $15.9$  & $11.0$      \\
\mlb{Pz27-0.5-10 \hfill  (A)} & $123$        & $75.6$        & $77.3$        & $117$  & $20.8$   & $993$    & $839$               & $-5.0$        & $16.4$  & $10.8$      \\
\mlb{Pz27-0.5-10 \hfill (B)} & $121$        & $74.7$        & $76.5$        & $117$  & $21.3$   & $1015$    & $842$               & $-5.1$        & $16.4$  & $10.4$      \\
\hline
 Mean of UEIS & $124(3)$        & $77(3)$        & $80(3)$        & $119(3)$  & $20.8(4)$   & $982(27)$    & $804(44)$               & $-5.2(1)$        & $16.3(3)$  & $11.0(6)$      \\
Literature \cite{Kiyono2016}        &  $117.64$      & $73.66$      & $73.46$  &  $110.17$      & $20.41$  & $959.10$    & $816.61$             & $-5.19$    & $16.06$& $11.59$      \\
Manufacturer  \cite{Meggitt2021}   & $147(4)$      & $105(3)$      & $94(2)$      & $113(3)$ & $23.0(6)$     & $1130(113)$ & $914(91)$            & $-3.1(2)$    & $16.0(8)$  & $11.6(6)$    \\
\hline
$\hat{\sigma}_\mr{UEIS} (\%)$  & \mrr{$1.1$}      & \mrr{$1.7$}      & \mrr{$1.8$}      & \mrr{$1.1$} & \mrr{$1.0$}     & \mrr{$1.4$} & \mrr{$2.8$}            & \mrr{$-1.5$}    & \mrr{$0.8$}  & \mrr{$2.6$}    \upspace    \\
$\Delta_\mr{UEIS}^\mr{lit} (\%)$  & \mrr{$-5.2$}\upspace      & \mrr{$-4.7$}      & \mrr{$-7.5$}      & \mrr{$-7.6$} & \mrr{$-1.8$}     & \mrr{$-2.3$} & \mrr{$1.6$}            & \mrr{$0.6$}    & \mrr{$-1.3$}  & \mrr{$5.5$} \\
$\Delta_\mr{UEIS}^\mr{manf} (\%)$  & \mrr{$18$}\upspace      & \mrr{$36$}      & \mrr{$18$}      & \mrr{$-5.2$} & \mrr{$11$}     & \mrr{$15$} & \mrr{$14$}            & \mrr{$-40$}    & \mrr{$-1.7$}  & \mrr{$5.5$}    \\
\end{tabular}
\end{ruledtabular}

\vspace*{-5mm}
\end{table*}

\begin{table*}[t]

\centering
\caption{\tablab{Pz27_fit_imag}
The imaginary part (double prime) of the material parameters $p_i^\mr{pz27}$ of the four Pz27 disks of \tabref{SampleDim} obtained by the UEIS method with initial values from the literature \cite{Kiyono2016} and the parameters $\ve''_{11}$, $e''_{31}$, $e''_{33}$, and $e''_{15}$ set to zero by assumption. The digits in brackets denote one standard deviation, $\hat{\sigma}_\mr{UEIS}$ is the relative standard deviation of the UEIS mean, and $\Delta_\mr{UEIS}^\mr{lit}$ is the relative deviation from UEIS of the values of Ref.~\cite{Kiyono2016} (lit).}
\begin{ruledtabular}
\begin{tabular}{c|cccccccccc}
Pz27 disk                                                     & $C''_{11}$  & $C''_{12}$  & $C''_{13}$  & $C''_{33}$& $C''_{44}$ & $\ve''_{11}$  & $\ve''_{33}$ & $e''_{31}$ & $e''_{33}$& $e''_{15}$  \\
               & (MPa)      & (MPa)      & (MPa) & (MPa)      & (MPa)      & ($\ve_0$) & ($\ve_0$)  & (C/m$^2$)  &(C/m$^2$) & (C/m$^2$)     \\
\hline
\mlb{Pz27-0.5-6.35 \hfill  (A)}\upspace  & $-709$        & $370$        & $121$        & $-280$  & $-582$   & $0$    & $3.9$               & $0$        & $0$  & $0$    \\
\mlb{Pz27-0.5-6.35 \hfill (B)} & $-628$        & $442$        & $122$        & $-377$  & $-596$   & $0$    & $2.5$               & $0$        & $0$  & $0$      \\
\mlb{Pz27-0.5-10 \hfill  (A)} & $-558$        & $448$        & $112$        & $-518$  & $-488$   & $0$    & $0.0$               & $0$        & $0$  & $0$      \\
\mlb{Pz27-0.5-10 \hfill (B)} & $-510$        & $510$        & $90.7$        & $-538$  & $-486$   & $0$    & $6.3$               & $0$        & $0$  & $0$      \\
\hline
Mean of UEIS & $-601(87)$        & $442(57)$        & $112(15)$        & $-428(122)$  & $-538(59)$   & $0$    & $3(3)$               & $0$        & $0$  & $0$      \\
Literature \cite{Kiyono2016}        &  $-460$      & $220$      & $110$  &  $-400$      & $-400$  & $-$    & $10.72$             & $-$    & $-$ & $-$   \\
\hline
$\hat{\sigma}_\mr{UEIS} (\%)$  & \mrr{$-7.2$}      & \mrr{$6.4$}      & \mrr{$6.6$}      & \mrr{$-14$} & \mrr{$-5.5$}     & $-$ & \mrr{$41$}            & $-$    & $-$  & $-$    \upspace \\
$\Delta_\mr{UEIS}^\mr{lit} (\%)$  & \mrr{$-23$}\upspace      & \mrr{$-50$}      & \mrr{$-1.4$}      & \mrr{$-6.6$} & \mrr{$-26$}     & $-$ & \mrr{$235$}            & $-$   & $-$  & $-$    \\
\end{tabular}
\end{ruledtabular}%
\end{table*}



\begin{figure}[t]
\centering
\includegraphics{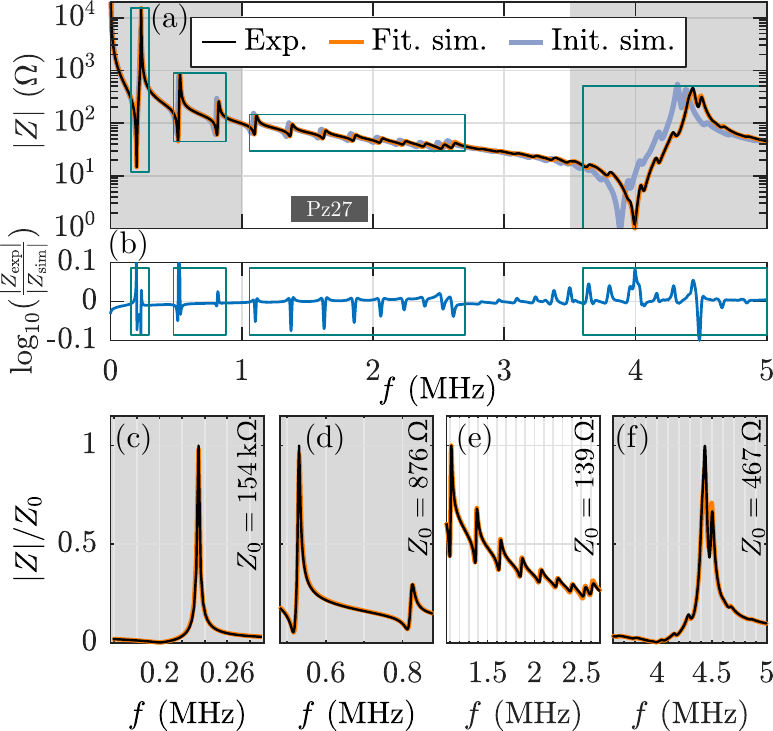}
\caption{\figlab{UEIS_Pz27}
(a) Semilog plot of the measured (black) and simulated UEIS $|Z(f)|$ of an unloaded  Pz27-0.5-10 disk. In the simulations are used the UEIS-fitted (orange) and initial literature (blue) Pz27 parameters listed in \tabsref{Pz27_fit_real}{Pz27_fit_imag}. The gray region indicates the frequency range used in the fitting. (b) The logarithmic difference $\Delta_\mr{sim}^\mr{exp} = \log_{10}(|Z_\mr{exp}|/|Z_\mr{sim}|)$ between measured and simulated impedance spectrum. (c)-(f) Zoom-in on different regions showing the measured and simulated spectrum on a linear and re-normalized scale. Each region is indicated by a frame in both (a) and (b).}
\end{figure}

\begin{figure}[t]
\centering
\includegraphics{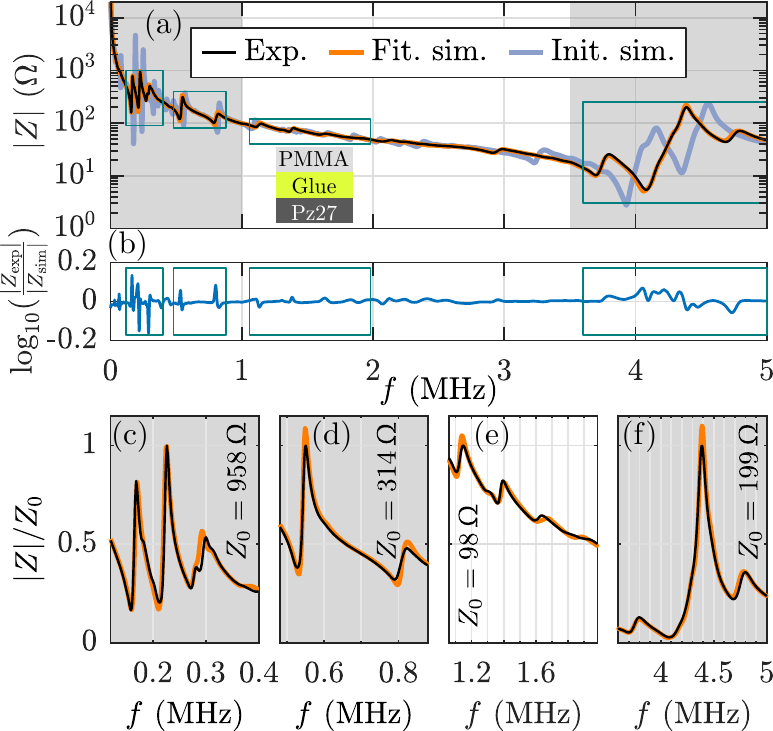}
\caption{\figlab{UEIS_PMMA}
(a) Semilog plot of the measured (black) and simulated UEIS $|Z(f)|$ of a PMMA-1.4-25 ring glued to a Pz27-0.5-10 disk by a 21-$\SImum$-thick layer of NOA 86H glue. The UEIS-fitted simulation (orange) is computed using the UEIS  parameter values listed in Tables~\tabnoref{Pz27_fit_real}, \tabnoref{Pz27_fit_imag}, \tabnoref{glue_fit}, and \tabnoref{PMMA_fit}. The initial-value simulation is shown in blue. The gray regions indicate the frequency ranges used in the fitting. (b) The logarithmic difference $\Delta_\mr{sim}^\mr{exp} = \log_{10}(|Z_\mr{exp}|/|Z_\mr{sim}|)$ between measured and simulated impedance spectrum. (c)-(f) Zoom-in on different regions showing the measured and simulated spectrum on a linear and re-normalized scale. Each region is indicated by a frame in both (a) and (b).}
\end{figure}

\subsection{UEIS-fitted material parameters for glue}
\seclab{fittedGlue}

The parameters $p_i^\mr{glue}$ of the used UV-cured NOA 86H glue were determined by the UEIS method as described in \secref{FitProc} using a UV-cured glue ring glued to a Pz27 disk with the dimensions listed in \tabsref{SampleDim}{config}. The resulting values for $C_{11}$ and $C_{44}$ are presented in \tabref{glue_fit} together with the corresponding values for the sound speeds $\cL$ and $\cT$, the attenuation coefficients $\aL$ and $\aT$, as well as Young's modulus $E$ and Poisson's ratio $\nu$. The expressions for these additional parameters, valid for any isotropic elastic material, are obtained by assuming frequency-independent moduli $C_{11}$ and $C_{44}$ in the limit of weak attenuation, $\big|C_{11}''\big| \ll C_{11}'$ and $\big|C_{44}''\big| \ll C_{44}'$, and by introducing the complex-valued wavenumbers $\kL = \omega/\cL +\ii\aL$ and $\kT = \omega/\cT +\ii\aT$,
 \bsuba{c_alf_E_nu}
 \balat{2}
 \cL &= \sqrt{\frac{C_{11}'}{\rho}},\qquad &
 \aL(f) &= \pi \sqrt{\frac{\rho}{\big(C_{11}'\big)^3}}\big|C_{11}''\big|\:f,
 \ealat
 \balat{2}
 \cT &= \sqrt{\frac{C_{44}'}{\rho}}, &
 \aT(f) &= \pi \sqrt{\frac{\rho}{\big(C_{44}'\big)^3}}\big|C_{44}''\big|\:f,
 \\
 \nu & = \frac{\frac12 C_{11}'- C_{44}'}{C_{11}'-C_{44}'} &
 E & = \frac{3C_{11}'-4C_{44}'}{C_{11}'-C_{44}'}\:C_{44}'.
 \ealat
 \esuba

\begin{table}[b]
\centering
\caption{\tablab{glue_fit}
The material parameters at $24~\SICel$ of the UV-cured NOA 86H glue determined by the UEIS method. The density of the glue was measured to be $\rho = 1.3(2)\times10^3~\SIkgm$, and the attenuation coefficients are for the center frequency $f_c = 1.90$~MHz of the UTT method described in \secref{UTTresults}.}
\begin{ruledtabular}
\begin{tabular}{lll}
Parameter & Parameter & Parameter \\
\hline
$C_{11}' = 4.65~\SIGPa$       &  $\cL = 1891~\SIm/\SIs$ & $E = 3.20~\SIGPa$\upspace \\
$C_{44}' = 1.21~\SIGPa$       &  $\cT = \;\; 965~\SIm/\SIs$ & $\nu = 0.32$\\
$C_{11}'' = -0.51~\SIGPa$ &  $\aL(f_c) = 346~\SINp/\SIm$ & \\
$C_{44}'' = -0.12~\SIGPa$ &  $\aT(f_c) = 613~\SINp/\SIm$ & \\
\end{tabular}
\end{ruledtabular}
\end{table}

\begin{table}[b]
\centering
\caption{\tablab{PMMA_fit}
The UEIS-fitted and UTT-measured elastic moduli for PMMA at $24~\SICel$ with a measured density $\rho = 1162(4)$~kg/m$^3$, and $\aL$ and $\aT$ evaluated at $1.90$~MHz. For UEIS, $\hat{\sigma}_\mr{UEIS}$ is the relative standard deviation of the UEIS mean. For UTT, the errors are based on $1\sigma$-prediction intervals around linear regression fits. $\Delta_\mr{UEIS}^\mr{UTT}$ is the deviation of the UTT values relative to the UEIS values.}
\begin{ruledtabular}
\begin{tabular}{cc|rd{3.2}rd{3.2}}
Param. & Unit & \mc{UEIS}   & \mrr{$\hat{\sigma}_\mr{UEIS}$ (\%)} & UTT  & \mrr{$\Delta_\mr{UEIS}^\mr{UTT}$ (\%)} \\
\hline
$C_{11}'$  & GPa & $7.18(4)$  & 0.6 & $7.1(1)$    & -1.1\upspace \\
$C_{11}''$ & GPa & $-0.183(5)$ & 2.9 & $-0.19(1)$  & 3.8     \\
$C_{44}'$  & GPa & $1.553(8)$ & 0.5 & $1.56(1)$   & 0.5      \\
$C_{44}''$ & GPa & $-0.111(7)$  & 6.3 & $-0.098(6)$      & -11.7     \\
$c_\mr{lo}$& m/s & $2486(8)$  & 0.3 & $2469(19)$  & -0.7     \\
$c_\mr{tr}$& m/s & $1156(4)$  & 0.3 & $1160(4)$   & 0.3      \\
$\alpha_\mr{lo}$ & Np/m & $61(2)$   & 3.2 & $66(4)$  & 8.2  \\
$\alpha_\mr{tr}$ & Np/m & $370(24)$ & 6.4 & $322(18)$ & -13.0   \\
$E$        & GPa & $4.23(2)$  & 0.5 & $4.25(3)$     & 0.5    \\
$\nu$      & --  & $0.362(1)$ & 0.4 & $0.358(3)$   & -1.1   \\
\end{tabular}
\end{ruledtabular}
\end{table}

\subsection{UEIS-fitted material parameters for PMMA}
\seclab{fittedPMMA}

With the characterization of the Pz27 transducer disk and the glue completed, we move on to the determination of the complex-valued elastic moduli $C_{11}$ and $C_{44}$ for PMMA, which in principle could have been any other elastic polymer. We studied four PMMA polymer rings with the dimensions listed in \tabref{SampleDim}, all around 1.4~mm thick and with diameters of 20 or 25~mm, and glued to Pz27 disks with the dimensions listed in \tabref{config}.

The resulting UEIS-fitted parameters $C_{11}'$, $C_{44}'$, $C_{11}''$, and $C_{44}''$ at 24~$\SICel$ for the PMMA  are listed in \tabref{PMMA_fit} together with the corresponding values obtained by the UTT technique. The relative standard deviation $\hat{\sigma}_\mr{UEIS}$ on the real parts is low ($\sim$0.5~\%), and an order of magnitude higher on the imaginary parts (3-6~\%). We find good agreement between the UEIS and the UTT values, in all cases with relative deviations $\Delta_\mr{UEIS}^\mr{UTT} < 3\hat{\sigma}_\mr{UEIS}$. In terms of the derived sound speeds, $\cL$ and $\cT$, and the derived Young's modulus $E$ and Poisson's ratio $\nu$, the relative deviation of UTT  values from UEIS values is around 0.5~\%. For the longitudinal and transverse attenuation $\aL$ and $\aT$ coefficients, the relative deviations of UTT values relative to UEIS values are higher, around 7-15~\%.

Again, likely due to the lower sensitivity of the $C_{11}''$ and $C_{44}''$ coefficients, it proves more difficult to obtain the imaginary parts of the elastic moduli than the real parts. Deviations of the UTT values from the UEIS values, may in part be explained by the fact that the UTT technique uses a frequency pulse with a width of 1~MHz around the center frequency 1.90~MHz, whereas UEIS is based on an entire frequency spectrum from 500~Hz to 5~MHz using a single frequency at a time. However, whereas different models exist, which assume a frequency-dependence of the elastic moduli of PMMA \cite{Ilg2012}, similar to the frequency dependencies measured in PDMS~\cite{Xu2020}, we do find it sufficient in the UEIS method to neglect the frequency-dependence of the complex-valued elastic moduli of PMMA.

In \figref{UEIS_PMMA} is shown an example of the measured and the simulated UEIS spectra for a PMMA ring glued to a Pz27 disk. We find a good agreement between the measured and the fitted simulated UEIS spectrum, and it can also be seen, how even smaller features of the experimental impedance curve are captured in the simulated frequency spectrum. Relative deviations $\Delta_\mr{sim}^\mr{exp} = \log_{10}\frac{|Z_\mr{exp}|}{|Z_\mr{sim}|}$ up to $20~\%$ are found in regions near resonance peaks. However, zoom-ins there show how a frequency-shift of a few percent can lead to high deviations $\Delta_\mr{sim}^\mr{exp}$, while still maintaining good agreement between measurement and simulation. For example, shifting a Lorentzian peak with a Q-value of $Q = 250$ by 1\% of its resonance frequency, results in a relative deviation $\sim10\%$.

We furthermore studied the impact of small deviations in the thickness and elastic moduli of the glue on the obtained coefficients for the polymer ring. A change of the Young's modulus $E_\mr{glue}$ by $\pm 1$~\% leads to changes in the real-valued coefficients $C_{11}'$ and $C_{44}'$ by less than 0.05~\%, while the $C_{11}''$ and $C_{44}''$ coefficients change by 0.4~\% and 0.6~\% respectively. In a separate numerical study, when changing the thickness of the glue layer from 12~$\SIum$ to 8~$\SIum$, a relative deviation of 33.3~\%, we observe a decrease in the real-valued coefficients $C_{11}'$ and $C_{44}'$ by 0.8~\% and 0.2~\% respectively. The relative changes for the imaginary-valued coefficients $C_{11}''$ and $C_{44}''$ are slightly higher by +2.4~\% and $-1.3$~\%, respectively, but still much lower when compared to the relative change in thickness and well in line with the identified uncertainties of the parameters listed in \tabref{PMMA_fit}.

As a further validation of the UEIS method, we use the UEIS-fitted values $p_i^\mr{pmma}$ to simulate selected resonance modes in the PMMA ring. Subsequently, as shown in Sec.~S5 in the Supplemental Material~\cite{Note1}, we have successfully compared these predicted modes with direct measurements of the corresponding modes obtained by using a single-point laser-Doppler vibrometer system VibroFlex Connect (Polytec, Waldbronn, Germany).

\section{Conclusion}
\seclab{conclusion}

We have developed a method based on measured and simulated ultrasound electrical impedance spectroscopy (UEIS) able to determine the frequency-independent complex-valued elastic moduli of polymers. The method is a two-step procedure: Firstly, the material parameters of the used, unloaded piezoelectric transducer disk are fitted by an inverse problem, matching the measured and simulated electrical impedance spectrum. Secondly, a polymer ring is glued onto the transducer, and the same technique is used to fit the complex-valued elastic moduli of the polymer. To evaluate its reproducibility, the method was applied on four different system geometries involving the polymer PMMA, achieving a relative error below $0.5\,\%$ for Young's modulus and Poisson's ratio, and below $7\,\%$ for the attenuation coefficients. The method was validated experimentally within the 3$\sigma$-level using ultrasonic through-transmission on PMMA samples.

It is noteworthy that the model assumption of fre\-quen\-cy-independent elastic moduli $C_{ik}$ leads to simulated UEIS spectra  $Z_\mr{sim}(f)$ that predicts the measured UEIS spectra $Z_\mr{exp}(f)$ so well in the entire frequency range from 500~Hz to 5~MHz as shown in \figsref{UEIS_Pz27}{UEIS_PMMA} for Pz27 and PMMA, as well as for the UV-curable glue NOA~86H (not shown). This frequency independence leads to the linear frequency dependence of the attenuation coefficients $\aL$ and $\aT$ exhibited in \eqref{c_alf_E_nu}, a linearity which can be contrasted with the $f^2$-dependence of $\alpha_0$ in Newtonian fluids, \eqref{alphaWater}, and the non-integer powers observed in typically softer materials, such as the $f^{1.456}$-dependence of $\aL$ and the $f^{0.924}$-dependence of $\aT$ observed in the rubber PDMS~\cite{Xu2020}. It is straightforward to include such frequency-dependency of the elastic moduli in the UEIS model, should materials with that property be studied. One simply modify the respective moduli and coupling coefficients in the constitutive equations \eqsnoref{ConstitPolymer}{ConstitPZT} at the cost of extending the list of parameters $p_i$ with the necessary parameters needed to describe the frequency dependency. For the relatively stiff polymer PMMA, the elastic modulus tensor $\CCC$ can be taken as frequency independent, whereas modeling softer, rubber-like materials, the frequency dependency of $\CCC$ must be taken into account \citep{Pritz1998}.

The UEIS technique extends the existing field of resonance ultrasound spectroscopy by making use of the electrical impedance spectrum over a wide frequency range of several MHz involving both on-resonance and off-resonance frequencies, it has no obvious frequency limitations before severe attenuation sets in above 100~MHz, and it contains information of all relevant parameters of the piezoelectric transducer disk, the glue layer, the polymer ring, and the geometry of the assembled stack. Experimentally, the technique is low-cost, easy-to-use, simple, and well-suited for materials used in ultrasound applications. The recording of a given impedance spectrum takes less than 4~minutes. Afterwards, within about 1~minute, the impedance spectrum can be loaded into our MATLAB script, and the automated UEIS fitting procedure is executed. After a run time of about 10 hours, the resulting UEIS-fitted impedance spectrum and the parameter values  $p_i^\mr{pz27}$, $p_i^\mr{glue}$, or  $p_i^\mr{pmma}$ are delivered by the software.

The UEIS technique is not limited to the chosen examples of Pz27, glue, and PMMA, but it can in principle be used on other classes of elastic materials including rubbers, glasses, and metals. We believe that the presented UEIS technique will become a valuable and easy-to-use tool in the ultrasound application fields mentioned in the introduction, by providing well-determined parameter values for the materials used, namely the relevant complex-valued elastic moduli at the relevant ultrasound frequencies.

\section{Acknowledgments}
We would like to thank Ola Jakobsson (Lund University) for setting up and introducing us to the laser-Doppler vibrometer system, Axel Tojo (Lund University) for his help with the UTT setup, Komeil Saeedabadi and Erik Hansen (DTU) for providing and cutting the polymer rings, and Erling Ringgaard (Meggitt) for useful discussions and for providing the Pz27 samples. This work is part of the Eureka Eurostars-2 E!113461 AcouPlast project funded by Innovation Fund Denmark, grant no.~9046-00127B, and Vinnova, Sweden's Innovation Agency, grant no.~2019-04500. Finally, the work was supported by Independent Research Fund Denmark, Technology and Production Sciences, grant no.~8022-00285B and the Swedish Foundation for Strategic Research, grant no.~FFL18-0122.


%

\end{document}